\documentstyle[preprint,eqsecnum,aps]{revtex}
\draft
\title{Alternative sets of hyperspherical harmonics: Satisfying cusp 
conditions through frame transformations}
\author{Thomas A. Heim\cite{TH} and Dmitry Green\cite{DG}} 
\address{James Franck Institute, University of Chicago, Chicago, IL 60637}
\begin{document}
%
%
\maketitle
\begin{abstract}
By extending the concept of Euler-angle rotations to more than three 
dimensions, we develop the systematics under rotations in higher-dimensional 
space for a novel set of hyperspherical harmonics. Applying 
this formalism, we determine all pairwise Coulomb interactions in a
few-body system without recourse to multipole expansions. Our approach
combines the advantages of relative coordinates with those of the
hyperspherical description. In the present method, each Coulomb matrix element
reduces to the ``$1/r$'' form familiar from the two-body problem. Consequently,
our calculation accounts for all the cusps in the wave function whenever an
inter-particle separation vanishes. Unlike a truncated multipole expansion, 
the calculation presented here is exact. Following the systematic
development of the procedure for an arbitrary number of particles, we 
demonstrate it explicitly with the simplest non-trivial example, the 
three-body system.
\end{abstract}
%
%
\section{Introduction}
A system consisting of $N$ charged particles gives rise to $N(N-1)/2$
pairwise Coulomb interaction terms in its Hamiltonian. 
Since only the two-body problem ($N=2$) can actually be solved
exactly, conventional atomic physics methods view the complete system of
particles at the outset as a conglomerate of independent two-body systems,
adding the interactions between these independent particles in a second step.
This approach amounts to selecting a suitable subset of the $N(N-1)/2$
Coulomb terms for which a solution in terms of ``simultaneous two-body
wave functions'' can be given. For instance, the independent-particle
model for atomic systems treats each electron as interacting primarily with
the nucleus (or with the ionic core in the case of valence electrons). Each
(valence) electron's position introduces an independent spherical direction.
Just as in the familiar solution of the hydrogen atom, each electron
contributes to the angular part of the total system's wave function a spherical
harmonic $Y_{lm}$ of the angles specifying its direction in space. In the next
step, all the electron-electron interactions are calculated by expanding the
corresponding separations into Legendre polynomials of the inter-electronic
angles, thus yielding the familiar multipole expansion. In more general terms,
the independent-particle model first selects a specific particle \#1 (on 
physical grounds, typically the nucleus or the ionic core) and solves for 
each of the remaining $N-1$ particles the two-body problem $\{1,j\}$,
$j=2,\ldots,N$, thereby providing a basis for expanding the total 
wave function. In the next step, the interaction between particles $i$ and
$j$ is calculated by adapting the reciprocal of their separation,
$1/r_{ij}$, to the coordinate system pertaining to the ``two-body'' basis 
functions for $i$ and $j$ ($i,j=2,\ldots,N$).  

The Coulomb interaction is singular whenever an inter-particle separation
vanishes. Since these singularities are isolated from one another, they 
do not pose fundamental difficulties in calculating the 
Hamiltonian matrix. However, they give rise to {\em cusps} (discontinuous 
derivatives) in the wave function \cite{Kato}, thus slowing down the
convergence of partial-wave expansions of the wave function in their vicinity 
\cite{Carroll,Hill}. The independent-particle wave functions can only account
(through the corresponding $s$-wave components) for the cusps arising from
vanishing separations $r_{1j}$, $j=2,\ldots,N$. Cusps due to vanishing 
$r_{ij}$ with $i\neq 1$ are not reproduced.
Possible remedies to these shortcomings include the explicit use of coordinates
like the inter-electronic angle $\theta_{ij}$ or even 
$r_{ij}$ besides $r_i$ and $r_j$; various approximations for $1/r_{ij}$ 
\cite{Bunge}; or replacing $r_i,r_j$ by 
$r_<=\min(r_i,r_j),r_>=\max(r_i,r_j)$ in the Hamiltonian
\cite{Goldman}. All these approaches amount   
to {\em adapting the interaction operator} $1/r_{ij}$ to a single coordinate
system.

The present investigation explores the alternative approach of {\em adapting 
the basis functions} to match the relevant inter-particle separations $r_{ij}$
by utilizing several coordinate systems simultaneously. We calculate a specific 
interaction term $\sim 1/r_{ij}$ in the coordinate system best suited for
this particular purpose. Thus, the particle separations $r_{ij}$ dictate the
choice of coordinate systems, the wave functions being transformed between
the relevant reference frames to evaluate the different terms of the 
Hamiltonian. The success of this approach hinges on our ability to
perform the numerous transformations between reference frames with high 
efficiency. Sec.~\ref{HSJacobi} describes hyperspherical Jacobi 
coordinates appropriate for this task. 
Sec.~\ref{algebra} provides the main results by implementing the relevant 
transformations for a system with an arbitrary number of particles and
by constructing basis functions (harmonics) suitable for extensive 
transformation between reference frames. 
The resulting set of hyperspherical harmonics, derived here in the context of
calculating the Hamiltonian matrix for a system of $N$ charged particles, has
in fact much wider applicability. 

Hyperspherical coordinates and corresponding hyperspherical harmonics have been
applied in various areas of physics since the 1950s, for instance, in
three-body scattering \cite{Feshbach,Wannier,Smith}, nuclear 
\cite{Delves,Smirnov} and
atomic \cite{Macek,Lin,Chen,Tang,Heim,Yang,Cavagnero,FanoRPP,FR86} physics, 
as well as in quantum chemistry \cite{Aquilanti,Avery}. However, 
the sets of functions introduced in the present investigation 
are equivalent to, but much
more flexible than, the hyperspherical harmonics discussed in 
\cite{Feshbach,Wannier,Smith,Delves,Smirnov,Macek,Lin,Chen,Tang,Heim,%
Yang,Cavagnero,FanoRPP,FR86,Aquilanti,Avery}.
Beyond constituting 
a complete orthogonal 
set of functions appropriate for expansions, their frame-independence
affords greater flexibility in analyzing selection rules and other relations
between harmonics. These aspects are conveniently investigated through ladder
operators; they are determined entirely by the symmetry properties of the 
$N$-particle system, independent of any coordinate representation. Only 
calculating the non-vanishing matrix elements requires an 
appropriate coordinate representation of the generic basis functions.
Sec.~\ref{example} illustrates the relevant procedures
with the simplest non-trivial example, the three-body problem. The concluding
Sec.~\ref{conclusion} discusses advantages and limitations of the new
technique, as compared to the conventional multipole expansion.

\section{Hyperspherical Jacobi coordinates} 
\label{HSJacobi}
The inverse proportionality between pairwise interaction and inter-particle
separation suggests replacing individual particles' positions with 
{\em relative} (Jacobi) coordinates.
Starting from one pair of particles, the construction of Jacobi coordinates
proceeds hierarchically by joining (the centers of mass of) increasingly
complex groups of particles. This hierarchical structure is
commonly referred to as a ``Jacobi tree'' \cite{Smirnov}. Alternative choices 
for the initial particle pair, as well as the ordering of successive particle 
groups, correspond to different Jacobi trees. In the next step, mass-scaling 
of the Jacobi coordinates fully exposes the symmetry of the kinetic 
energy operator for the multi-particle system. Let $p,q$ denote two (groups of) 
particles with masses $M_p, M_q$ and center of mass positions 
$\bbox{r}_p$ and $\bbox{r}_q$, respectively.
Appropriate mass-scaling of the relative coordinate in the form \cite{Smith}
$\bbox{\xi}_{p,q}=\{M_pM_q/(M_p+M_q)\}^{1/2}(\bbox{r}_p-\bbox{r}_q)$ removes
the individual mass dependence from the expression for the kinetic energy:
\begin{equation}
-\sum_{i=1}^N\frac{\hbar^2}{2M_i}\Delta_{\bbox{r}_i} = -\frac{\hbar^2}{2}
\sum_{k=1}^{N-1}\Delta_{\bbox{\xi}_k}-\frac{\hbar^2}{2M_{\rm tot}}
\Delta_{\bbox{r}_{\rm CM}}.
\end{equation}
Setting the origin at the center of mass of the whole system allows 
discarding the CM's position and motion. With the individual mass factors
removed, the kinetic energy (generalized Laplacian) displays complete symmetry
under rotations in $(3N-3)$-dimensional space.

Hyperspherical coordinates exploit this symmetry by
separating the ``shape'' of the system (described by $3N-4$ angular 
coordinates) from its overall ``size'' $R^2=\sum_k\bbox{\xi}_k^2 $ (with the 
dimension of a moment of inertia). Eigenfunctions of the Laplacian's angular 
part, termed ``hyperspherical harmonics'' by generalizing the two- and
three-dimensional cases, constitute a basis for 
expanding the complete
wave function. In higher-dimensional spaces, the angular Laplacian's eigenvalues
$\Lambda(\Lambda+3N-5)$ are highly degenerate. In addition to the ``grand
angular momentum'' $\Lambda$ \cite{Smith}, the 
hyperspherical harmonics thus require numerous labels---analogues of the 
single ``magnetic'' quantum number $m$ for three-dimensional spherical 
harmonics---for their
unique specification. Different sets of hyperspherical harmonics, 
resulting from solving the Laplacian eigenvalue problem by
separation of variables in alternative sets of coordinates, have been
investigated extensively (see, e.g., \cite{Smirnov,Cavagnero,Avery} for 
systematic studies). However, their
construction through separation of variables inevitably ties these harmonics
to a specific coordinate system. They are thus not well suited for extensive
transformation between different reference frames. The definition of
harmonics based only on their behavior under the relevant transformations is
the main result of this paper, to be derived in Sec.~\ref{algebra}.

The construction of each Jacobi tree starts with a pair of particles.
Thus, each Jacobi tree contains at least one Jacobi vector joining two
particles only (rather than centers of mass of particle {\em groups}).
We denote such a vector as a ``primary'' coordinate.
In any one Jacobi tree, up to $\lfloor N/2\rfloor$ Jacobi vectors are directly
proportional to actual inter-particle separations, $\bbox{r}_{ij}$, 
the remaining relative coordinates
necessarily involving larger complexes of particles. (Here and in the
following, $\lfloor x\rfloor$ denotes the largest integer not exceeding $x$.)
We will calculate each of the $N(N-1)/2$ Coulomb terms of the Hamiltonian
in a Jacobi tree where it occurs as a ``primary'' coordinate.
Obviously, all these terms then take the same form as the simple one-electron
integral over $1/r$ in hydrogen, but the variables now refer to a multitude of
different reference frames. Having thus eliminated all nested integrations,
our next task consists in  determining the transformations between different
Jacobi trees.

The most general transformation from one Jacobi tree to another (for the same
system of particles) resolves into a sequence of elementary operations.
Each elementary step consists in ``transplanting'' a sub-complex
$q$ from some particle complex $\{pq\}$ to a complex $\{qr\}$ \cite{Smirnov};
it is achieved by a two-dimensional kinematic rotation \cite{Smith} 
through the angle 
\begin{equation}
\phi=\tan^{-1}\sqrt{M_q(M_p+M_q+M_r)/M_pM_r}
\end{equation}
in the $(\bbox{\xi}_{p,q},\bbox{\xi}_{pq,r})$-plane  of the $(N-1)$-dimensional
space of mass-scaled Jacobi vectors:
\begin{mathletters}
\begin{eqnarray}
\bbox{\xi}_{q,r}' & = & \cos\phi\;\bbox{\xi}_{p,q}-\sin\phi\;
\bbox{\xi}_{pq,r} \\
\bbox{\xi}_{p,qr}' & = & \sin\phi\;\bbox{\xi}_{p,q}+\cos\phi\;\bbox{\xi}_{pq,r}.
\end{eqnarray}
\end{mathletters}
As each Jacobi vector is a vector in three-dimensional physical space, the 
basic rotation in the $(\bbox{\xi}_j,\bbox{\xi}_k)$-plane implies three 
rotations
through the same angle $\phi$ in the $(x_j,x_k)$-, the $(y_j,y_k)$-, and the
$(z_j,z_k)$-planes, where $(x,y,z)$ denote the Cartesian components of 
$\bbox{\xi}$. An elementary kinematic rotation through a finite angle $\phi$ in
the $(\bbox{\xi}_j,\bbox{\xi}_k)$-plane then reads \cite{Rau}
\begin{equation}\label{finiterot}
T_{\bbox{\xi}_j,\bbox{\xi}_k}(\phi)=
\exp(i\phi J_{x_jx_k})\,\exp(i\phi J_{y_jy_k})\,\exp(i\phi J_{z_jz_k}),
\end{equation}
with infinitesimal rotation operators 
\begin{eqnarray} 
J_{uv} & \equiv  & -i\left(u\frac{\partial}{
\partial v}-v\frac{\partial}{\partial u}\right), 
%
%
\qquad
(u,v)=(x_j,x_k),\;(y_j,y_k),\;(z_j,z_k). 
\label{infrotop}  
\end{eqnarray}
(For a heuristic explanation of (\ref{finiterot}),
recall the Taylor expansion of $f(x+\Delta x)$ which may formally be written
as $\exp(i\Delta x\, p)\,f(x)$ with the infinitesimal ``translation operator''
$p\equiv -i\frac{\partial}{\partial x}$ corresponding to the quantum 
mechanical linear momentum (with $\hbar\equiv1$).
In the present context, the translation from $x$ to $x+\Delta x$ is replaced
by a rotation through an angle $\phi$ from some direction in multi-dimensional
space to a new direction.) In order to determine 
$T_{\bbox{\xi}_j,\bbox{\xi}_k}$'s effect on the basis functions, we need to (i) 
express the generic 
rotation operators $J_{uv}$ in terms of ``ladder operators'' whose action on
the basis functions is straightforward to calculate, and (ii) 
construct basis functions suitable for such transformations. 

This section concludes with the outline of a strategy for determining the most 
efficient sequence of two-dimensional rotations to achieve the transformation
between arbitrary Jacobi trees. Note first that the individual terms in the 
Hamiltonian  are referred to by means of {\em particle labels}, implying the
need for appropriate antisymmetrization of the total wave function under 
interchange of identical particles. This consideration dominates the 
construction of the first Jacobi tree to which all subsequent transformations
are applied. In an atomic or molecular system, antisymmetrization 
concerns primarily the electronic part of the wave function. Thus the basic
Jacobi tree starts out with a pair of electrons. It grows by adding one 
electron at a time, finishing with the addition of nuclei or ionic core(s).
This procedure results in a ``canonical'' Jacobi tree \cite{Smirnov},
represented by the sequence of particle labels
\begin{equation}
(\cdots((((12)3)4)5)\ldots),
\end{equation}
whose parentheses separate sub-complexes. Alternatively, one could start 
by first forming as many {\em pairs} of particles as possible, i.e., 
$\lfloor N/2\rfloor$ pairs, before joining these pairs into larger 
complexes. Although the antisymmetrization of an electron pair is particularly
compact (namely, their relative angular momentum and their spin must add to an
{\em even} value), antisymmetrization among different pairs requires breaking 
up all these pairs, in essence going back to a canonical tree. Since breaking 
up sub-complexes involves about as many elementary rotations as forming the 
new complexes, this procedure is not efficient. Nevertheless, a set of $N$ 
(for odd particle number) or $N-1$ (even particle number) trees with the 
maximum number of $\lfloor N/2\rfloor$ pairs suffices to isolate all the 
$N(N-1)/2$ inter-particle separations. A successful strategy therefore aims at 
building these particular Jacobi trees from the canonical tree. The simplest 
of these ``pair-trees'', $(\cdots((12)(34))(56)\ldots)$, is obtained from the 
canonical tree with only $\lfloor N/2\rfloor-1$ rotations. In general,
the pair $(jk)$ with $j<k$ is $(k-j-\delta_{j,1})$ rotations away from the 
canonical tree, but many other ``useful''
pairs are formed in the course of these ``transplantations''.
\section{Transforming the basis functions}
\label{algebra}
This section provides the tools required to rotate harmonics from one
coordinate system to another. Our transformation matrices are analogous 
to the so-called Raynal-Revai coefficients \cite{Novoselsky}.
The reader familiar with Lie algebra will of course 
recognize the relevant aspects of $so(3N-3)$, the algebra of rotations in
$(3N-3)$ dimensional space. However, such familiarity is not presumed, and
we hardly use the terminology of group theory. 
Our presentation reflects a more pragmatic point of view,
emphasizing both the technical implementation as well as its relation to the
physical application at hand rather than full mathematical generality. 
For the latter aspects, the reader should turn to
the mathematical literature \cite{group}. 
\subsection{Rotations in $d$-dimensional space}
The most intuitive description of a rotation in three dimensions requires
two elements: the (invariant) {\em axis}, and the {\em angle} of rotation.  
The rotation itself occurs in a {\em plane} perpendicular to the axis of
rotation. In three dimensions, specifying the direction of the axis of
rotation is equivalent to, but more economical than, describing the actual 
plane of rotation. 
For a higher dimensional space this is no longer the case, because there
are several invariant directions perpendicular to a given plane. Thus, 
in $d$-dimensional space, a basic rotation is more appropriately characterized
as occurring in the 
{\em plane} spanned by two coordinates rather than by an invariant axis
orthogonal to it. The generic infinitesimal rotation operators then
take the form (\ref{infrotop}).
The number of different planes, $\frac{1}{2}d(d-1)$
in $d$ dimensions, equals the number of basic rotations. Because rotations
occurring in non-intersecting planes affect different pairs of coordinates, 
they are evidently independent of each other; the
corresponding rotation operators commute with one another. As there are
$\lfloor d/2\rfloor$
non-intersecting planes in a $d$-dimensional space, the largest set of
simultaneously commuting operators $J_{uv}$ 
contains $\ell\equiv \lfloor d/2\rfloor$ elements. These commuting
operators are commonly denoted as $H_j,\;j=1,\ldots,\ell$ \cite{group}. 
For our application,
the power of the Lie algebraic method derives from the
structure and the properties of rotations being completely independent of any 
particular coordinate representation of the operators $H_j$. In fact, 
the following manipulations are most efficiently carried out in generic 
Cartesian coordinates without any reference to a specific Jacobi tree. 

Simultaneous eigenfunctions of the $H_j$ with (integer) eigenvalues $m_j$,
$j=1,\ldots,\ell$, provide a basis for building hyperspherical harmonics 
suitable
for extensive rotations between reference frames. If the infinitesimal
rotation corresponding to $H_j$ occurs in the $(u_j,v_j)$-plane, the
simultaneous eigenfunction of the $\ell$ $H_j$ with eigenvalues $m_j$,
respectively,  reads 
\begin{equation}\label{efHi}
F(\{u_k,v_k\})\times\prod_{j=1}^{\ell}(u_j+i\, v_j)^{m_j} 
\end{equation}
with a function $F$ that bears further specification in
Subsect.~\ref{harmconst}. At this point, we only require that $H_j\,F\equiv 0$ 
for all $j$.
The set of eigenvalues
$\bbox{\mu}=\{m_1,m_2,\ldots,m_{\ell}\}$ serves as a label identifying 
different harmonics with the same value of $\Lambda(\Lambda+d-2)$ in the
angular Laplacian's eigenvalue equation. Additional labels required for a 
unique specification will be introduced in Subsect.~\ref{harmconst}.

Appropriate linear combinations of the remaining infinitesimal rotation
operators act as {\em ladder operators} $E_{\bbox{\alpha}}$ satisfying
\begin{equation}  
[H_j,E_{\bbox{\alpha}}]=\alpha_j\, E_{\bbox{\alpha}},\qquad j=1,\ldots,\ell,
\end{equation}
with an $\ell$-dimensional vector index $\bbox{\alpha}$ with components 
$\alpha_j=0,\pm 1$. Here $\alpha_j=0$ indicates that $E_{\bbox{\alpha}}$ 
does not change the part of the eigenfunction pertaining to $H_j$, whereas 
$\alpha_j=\pm 1$ means that it maps this part to the eigenfunction with
eigenvalue $m_j\pm1$. The ladder operators interrelate eigenfunctions with
different $\bbox{\mu}$ but degenerate $\Lambda$.
Raising and lowering operators form Hermitian conjugate pairs 
$E_{-\bbox{\alpha}}=E_{\bbox{\alpha}}^\dag$.

In general (for $\ell > 1$), each ladder operator affects
two of the $m_j$ simultaneously; for odd-dimensional spaces 
a subset of $\ell$ pairs of raising and lowering operators change one 
of the $m_j$ only \cite{group}. 
Abbreviating the vector label $\bbox{\alpha}$ by giving 
only its two non-vanishing components $\alpha_j$ and $\alpha_k$, the ladder 
operator that raises $m_j$ and simultaneously lowers $m_k$ takes the form
\begin{mathletters}
\begin{eqnarray}
E_{jk}^{+-}&=&-\frac{i}{2}\left((u_j+i\,v_j)
\frac{\partial}{\partial(u_k+i\,v_k)}
%
%
-(u_k-i\,v_k)\frac{\partial}{\partial(u_j-i\,v_j)}\right),\label{updown}
\end{eqnarray}
whereas the operator raising both $m_j$ and $m_k$ 
is represented by
\begin{eqnarray}
E_{jk}^{++}&=&-\frac{i}{2}\left((u_j+i\,v_j)
\frac{\partial}{\partial(u_k-i\,v_k)}
%
%
-(u_k+i\,v_k)\frac{\partial}{\partial(u_j-i\,v_j)}\right).\label{upup}
\end{eqnarray}
(Straightforward application of these operators to eigenfunctions 
(\ref{efHi}) verifies their behaving as ladder operators: they contribute
to, or remove from, (\ref{efHi}) factors $(u_j\pm iv_j)$ and 
$(u_k\pm i v_k)$ as appropriate for the intended ``ladder operator action''.)
For odd-dimensional spaces, a residual coordinate, denoted here by $w$, does
not occur in any of the $H_j$. The ladder operators changing only 
$m_j$ (rather than a pair $m_j,m_k$) read then 
\begin{equation}\label{Lpm}
E_j^{\pm}=-i\left((u_j\pm i\,v_j)\frac{\partial}{\partial w}-
w\frac{\partial}{\partial (u_j\mp iv_j)}\right),\qquad j=1,\ldots,\ell.
\end{equation}
\end{mathletters}
(For $d=3$, setting $(u,v,w)=(x,y,z)$ and transforming the derivatives reveals
the familiar pair of ladder operators $l_x\pm i\,l_y$, the single $H_j$ 
occurring in this case coinciding with $l_z$.)
This symbolic representation allows for efficient implementation on the
computer. 

A subset of $\ell$ ladder operators $E_{\bbox{\epsilon}_j}$ 
(and their Hermitian conjugates) suffices to interrelate all harmonics with 
the same eigenvalue $\Lambda$. A convenient choice \cite{group} for
these $E_{\bbox{\epsilon}_j}$ is given by the ladder operators that raise 
$m_j$ and simultaneously lower $m_{j+1}$ for $j=1,\ldots,\ell-1$.
$E_{\bbox{\epsilon}_{\ell}}$'s form depends on whether $d$ is even or odd.
For $d$ even, $E_{\bbox{\epsilon}_{\ell}}$ raises both $m_{\ell-1}$ and 
$m_\ell$, and for odd $d$, it raises $m_\ell$ only, without changing
any of the other $m_j$. The set $\{\bbox{\epsilon}_j\}$, $j=1,\ldots,\ell$,
 then provides a basis 
for the
$\ell$-dimensional space of the vector labels $\bbox{\alpha}$ and $\bbox{\mu}$.
In an $\ell$-component vector notation, this basis reads (we assume
$\ell\ge5$ in order to expose the generic structure clearly)
\begin{equation}
\begin{array}{rcl}
\bbox{\epsilon}_1 & = & (1,-1,0,0,\ldots,0) \\
\bbox{\epsilon}_2 & = & (0,1,-1,0,\ldots,0) \\
\bbox{\epsilon}_3 & = & (0,0,1,-1,\ldots,0) \\
 & \vdots & \\
\bbox{\epsilon}_{\ell-1} & = & (0,0,\ldots,0,1,-1) \\
\bbox{\epsilon}_{\ell} & = & \left\{ \begin{array}{ll} (0,0,\ldots,0,1,1) &
\quad \text{for } d=2\ell \\
(0,0,\ldots,0,0,1) & \quad\text{for } d=2\ell+1. \end{array}\right.
\end{array}
\end{equation}
In general these basis vectors are {\em not} orthogonal in $\ell$-dimensional 
space. Note, however, the following special cases:
(i) In four-dimensional space (with $\ell=2$), the two basis vectors 
 $\bbox{\epsilon}_1=(1,-1)$ and $\bbox{\epsilon}_2=(1,1)$ {\em are
 orthogonal}, indicating that the two basic ladder operator
 pairs, $E_{\pm\bbox{\epsilon}_1}$ and $E_{\pm\bbox{\epsilon}_2}$,
 commute with each other. In 
 group theoretical language, this feature reflects the direct product 
 structure $SO(4)=SO(3)\otimes SO(3)$. However, the two $SO(3)$-components do 
 not refer to $m_1$ and $m_2$ directly, but rather to $(m_1-m_2)/2$ and 
 $(m_1+m_2)/2$. 
(ii)  For rotations in three-dimensional space, we have $\ell=1$, thus only one
 quantum number $m$ which is being changed by one pair of ladder operators 
 $E_{\pm\bbox{\epsilon}_1}\equiv l_{\pm}$. 
(iii) In $d=2$ dimensions, {\em there are no ladder operators}. 
Since all rotations
occur in the same plane (the ``only'' plane of two-dimensional space), 
the order in 
which rotations through different angles are performed does not matter;
they are all independent of each other. 
In the present formulation, the (single) rotation operator $H_1$ generates all 
rotations. 
 Each rotation is associated with its own harmonic function, $\exp(i\phi)$, the
 phase functions for different rotations (i.e., for different rotation angles
 $\phi$)  not being related to one another through linear operators. 
\subsection{Rotation of hyperspherical harmonics}
\label{rotharms}
We now turn to the analysis of 
the transformation described by (\ref{finiterot}).
Note first that the three rotations occur in three non-intersecting planes,
affording the more suitable representation 
$\exp(i\phi[J_{x_jx_k}+J_{y_jy_k}])\exp(i\phi\,J_{z_jz_k})$. 
Arranging the Cartesian components of the $N-1$
mass-scaled Jacobi vectors $\bbox{\xi}_k$ in the form
\begin{equation} \label{arrangement}
\{x_1,y_1,x_2,y_2,\ldots,x_{N-1},y_{N-1},z_1,z_2,\ldots,z_{N-1}\},
\end{equation}
we choose the $H_j$ by selecting pairs of coordinates from this list,
starting from the left:
\begin{mathletters}
\begin{eqnarray}
H_j & = & J_{x_jy_j}, \qquad j=1,\ldots,N-1 \label{Hxy} \\
H_{N-1+k} & = & J_{z_{2k-1}z_{2k}}, \qquad k=1,\ldots,\left\lfloor\frac{N-1}{
2}\right\rfloor,\label{Hzz}
\end{eqnarray}
\end{mathletters}
in the notation of (\ref{infrotop}).
We have stressed repeatedly our aim of defining hyperspherical harmonics
entirely through their behavior when acted upon by the operators
$H_j$ and $E_{\bbox{\epsilon}_j}$, because these operators embody the 
structure and symmetry of the whole system in a frame-independent way. 
The explicit coordinate representation (\ref{Hxy}--\ref{Hzz}) of the relevant 
operators appears at variance 
with our intended frame-independence. However, the spatial coordinates,
$(x_j,y_j,z_j)$, $j=1,\ldots,N-1$, in (\ref{Hxy}--\ref{Hzz}) should be viewed
as {\em generic Cartesian coordinates}; they are {\em not} tied to any 
particular Jacobi tree. Arranging the generic Cartesian coordinates as we 
did in (\ref{arrangement}), on the other hand, does reflect physical
considerations beyond the purely mathematical structure of rotations in
higher-dimensional space: The latter would instead label the coordinates
most appropriately as $(X_1,X_{-1},X_2,X_{-2},\ldots,X_\ell,X_{-\ell})$,
supplemented possibly with an $X_0$ for odd-dimensional spaces. Our
arrangement takes into account that the dimension $(3N-3)$ arises from a
{\em product structure} of an $(N-1)$-dimensional particle space with the
three-dimensional physical space of each Jacobi vector. In particular, the
arrangement (\ref{arrangement}) affords attributing {\em relevant physical
meaning} to the first $(N-1)$ eigenvalues $m_j$, $(j=1,\ldots,N-1)$: They
represent the $z$-projections of physical angular momenta; their sum 
constitutes the $z$-projection $L_z$ of the total orbital angular momentum, 
an invariant of the system.

As is evident from (\ref{Hxy}), the coordinates 
$(x_j,x_k,y_j,y_k)$ occur in $H_j$ and $H_k$. The $x$- and $y$-parts of the 
rotation 
(\ref{finiterot}) thus involve the ladder operators that change $m_j$ and
$m_k$ only. A more detailed analysis shows that (\ref{updown}--c) can be 
inverted to read:
\begin{equation}
J_{x_jx_k}+J_{y_jy_k} = E_{jk}^{+-}+E_{jk}^{-+}
\equiv E_{\bbox{\alpha}(jk)}+
E_{-\bbox{\alpha}(jk)}, \label{rotxy}
\end{equation}
for $1\le j < k \le N-1$,
with $\bbox{\alpha}(jk)=\sum_{s=j}^{k-1}\bbox{\epsilon}_s$, i.e., the
$\ell$-component vector with $+1$ as its $j$-th component and $-1$ as
its $k$-th component, all other entries being 0. 

The following consideration is central to our development, providing the
crucial link between the ladder operator representation (\ref{rotxy}) for
$(J_{x_jx_k}+J_{y_jy_k})$ and their transformation matrix elements, by means
of Euler-angle rotations. In analogy to the relation 
$l_x=\frac{1}{2}(l_++l_-)$ familiar from rotations in
three dimensions, we view the sum of a ladder operator $E_{\bbox{\alpha}(jk)}$ 
and its inverse (Hermitian conjugate) $E_{-\bbox{\alpha}(jk)}$ as describing a 
rotation about an analogue of the $x$-axis \cite{A1}. 
In three dimensions, an arbitrary rotation conventionally
resolves into a sequence of three rotations: about the $z$-axis, about the 
resulting $y'$-axis, and about the new $z'$-axis, through the Euler angles 
$(\varphi,\theta,\psi)$, respectively \cite{Rau}. In terms of these three
Euler-angle rotations, a rotation about the $x$-axis
through an angle $\phi$ results from the following sequence of operations:
The first Euler-angle rotation about the $z$-axis through the angle 
$\varphi=-\pi/2$ rotates the $y$-axis onto the original $x$-axis; in the
second step one rotates about this $y'$-axis (which is the original $x$-axis) 
through the angle $\theta=\phi$; the third Euler-angle rotation finally 
moves the $x'$-axis back to the original $x$-direction by rotating about 
the $z'$-axis (lying in the original $(yz)$-plane) through the angle 
$\psi=\pi/2$.
Wigner's $d$-symbol is the matrix element for the rotation of a spherical 
harmonic about the $y$-axis,  
$d_{m'm}^{(l)}(\phi)= \langle Y_{lm'}|\exp(i\phi l_y)|Y_{lm}\rangle$, 
whereas the two $z$-type rotations only
contribute phase factors $\exp(i\,m'\pi/2-i\,m\pi/2)$ \cite{Rau}, giving 
for the rotation of a spherical harmonic about the $x$-axis
\begin{equation} \label{rotx3d}
\langle Y_{lm'}|\exp(i\phi l_x)|Y_{lm}\rangle= e^{im'\frac{\pi}{2}}\;
d_{m'm}^{(l)}(\phi)\;e^{-im\frac{\pi}{2}}.
\end{equation}

Generalization of the concept of Euler-angle rotations from three to more 
dimensions hinges on the following key observations: $m$ and $m'$ serve 
to distinguish between degenerate harmonics with the same $l$. 
In higher-dimensional spaces, the vector indices $\bbox{\mu}$ and $\bbox{\mu}'$
play the same role. However, the parameter $l$ in the 
$d$-symbol indicates not only the angular Laplacian's eigenvalue $l(l+1)$, 
but also---more importantly---the range of possible $m$ values, 
$-l\le (m,m')\le l$. More precisely, it sets the upper limit of $2l$ for
the number of times either one of the two ladder operators $l_+,l_-$ can act 
in direct succession before necessarily mapping any spherical harmonic
$Y_{lm}$ to zero. 
(Since $l_+Y_{lm}\mapsto Y_{l,m+1}$, it takes $l-m\le 2l$
``raising operator'' steps to reach $Y_{ll}$ from $Y_{lm}$. 
Applying $l_+$ in succession 
$(2l+1)$ times necessarily raises $m$ beyond its upper limit.
The absence of a corresponding harmonic 
translates to $(l_+)^{2l+1}Y_{lm}\equiv 0$.). 
Viewed in this way, the harmonic's parameter $m$ indicates
its ``position'' along the string (of length $2l+1$) of degenerate harmonics 
interrelated by a ladder operator. We now extend these concepts 
to rotations in more than three dimensions.

The single $m$ ($l_z$'s eigenvalue) in three dimensions belongs to a 
string ranging from $-l$ to $+l$, accessed by the single pair of ladder 
operators $l_+,l_-$. In more than three dimensions, we replace it with an 
$\ell$-component vector $\bbox{\mu}$ whose components are changed (typically 
in pairs) along
different strings labeled by corresponding ladder operator pairs
$E_{\bbox{\alpha}}$ and $E_{-\bbox{\alpha}}$. (Recall, however, that the 
$\ell$ vectors $\bbox{\epsilon}_j$
provide a basis for the vectors $\bbox{\alpha}$; thus any ladder operator 
$E_{\bbox{\alpha}}$ can be expressed as products of $E_{\bbox{\epsilon}_j}$.)
The role of $m$ as an indicator of the harmonic's position along the single 
string in three dimensions extends therefore to higher dimensions if we project
the ``indicator'' $\bbox{\mu}$ onto the ``direction'' of any particular string
$\bbox{\alpha}$ of interest, i.e., if we define
\begin{equation}\label{muproj}
m_{\bbox{\alpha}} = \frac{\bbox{\mu}\cdot\bbox{\alpha}}{\bbox{\alpha}\cdot
\bbox{\alpha}},\qquad
m'_{\bbox{\alpha}} = \frac{\bbox{\mu}'\cdot\bbox{\alpha}}{\bbox{\alpha}\cdot
\bbox{\alpha}},
\end{equation}
where the scalar product in the denominators accounts for $\bbox{\alpha}$'s
{\em two} non-zero components in the case of a pairwise change of $m_j$'s.
In analogy to $l$ in the three-dimensional case, a parameter 
$\lambda_{\bbox{\alpha}}$ determines how many times the ladder operators 
$E_{\pm\bbox{\alpha}}$ can be applied in direct succession. 
The modulus of $m_{\bbox{\alpha}}$ in (\ref{muproj}) provides a 
lower limit for the relevant ``string length'' $\lambda_{\bbox{\alpha}}$. 
Since the $\bbox{\epsilon}_j$ form a basis for the $\bbox{\alpha}$, 
we anticipate that $\lambda_{\bbox{\alpha}}$ emerges from an 
$\ell$-component vector $\bbox{\lambda}=(\lambda_{\bbox{\epsilon}_1},\ldots,
\lambda_{\bbox{\epsilon}_{\ell}})$. The latter provides the additional
parameters required for a unique specification of harmonics (besides 
$\bbox{\mu}$ and the eigenvalue $\Lambda$) . Note 
that  in the general case  $\lambda_{\bbox{\alpha}}$ is not identical with 
the eigenvalue $\Lambda$, at variance with the situation in three dimensions.

To summarize the procedure so far, we note that the $x$- and $y$-parts of the 
finite rotation (\ref{finiterot}) affect only the two components $m_j$ and
$m_k$ of the vector $\bbox{\mu}$, shifting $m_j$ by an integer amount $n$ while
simultaneously changing $m_k$ by the same amount in the opposite direction. 
The projections (\ref{muproj}) evaluate to
\begin{mathletters}
\begin{eqnarray}
m_{\bbox{\alpha}(jk)}&=&\frac{1}{2}(m_j-m_k);\\
m_{\bbox{\alpha}(jk)}'& = & 
\frac{1}{2}(m_j'-m_k')=m_{\bbox{\alpha}(jk)}+n,
\end{eqnarray}
\end{mathletters}
with non-vanishing matrix elements for the rotation of a hyperspherical
harmonic $Y_{\Lambda,\bbox{\mu},\bbox{\lambda}}$ into another harmonic 
$Y_{\Lambda',\bbox{\mu}',\bbox{\lambda}'}$ occurring only if 
$\bbox{\mu}$ and $\bbox{\mu}'$ lie along the {\em same} $\bbox{\alpha}$-string 
of harmonics with degenerate $\Lambda$, i.e., if an integral
number $n$ of ladder operator steps $E_{\pm\bbox{\alpha}(jk)}$ separates
$\bbox{\mu}'$ from $\bbox{\mu}$. With our previous considerations on 
Euler-angle rotations leading to (\ref{rotx3d}), this matrix element reads 
%
%
\begin{equation}
\langle Y_{\Lambda',\bbox{\mu}',\bbox{\lambda}'}|\exp(i\phi[J_{x_jx_k}+J_{
y_jy_k}])| Y_{\Lambda,\bbox{\mu},\bbox{\lambda}}\rangle = 
%
%
e^{i(m'_{\bbox{\alpha}}-m^{}_{\bbox{\alpha}})\pi/2}\,  
d_{m'_{\bbox{\alpha}}m^{}_{\bbox{\alpha}}}^{(\lambda_{\bbox{\alpha}})}(2\phi)\,
\delta_{\Lambda',\Lambda}\delta_{\bbox{\lambda}',\bbox{\lambda}}
\delta_{\bbox{\mu}',\bbox{\mu}+n\bbox{\alpha}},\label{xypart}
\end{equation}
%
%
where we omitted $\bbox{\alpha}$'s parameters $j,k$ for brevity. The Kronecker 
symbols $\delta$ ensure that only harmonics with the same eigenvalue
$\Lambda$, lying along the same $\bbox{\alpha}$-string, 
are connected. The 
factor 2 multiplying the rotation angle $\phi$ reflects the absence of the
factor $1/2$ in the RHS of (\ref{rotxy}), as compared to the expression
for $l_x=\frac{1}{2}(l_++l_-)$, thereby effectively doubling the rotation
angle in (\ref{rotx3d}).

It remains to apply the same concepts to the $z$-part of the rotation 
(\ref{finiterot}). This part affects only 
$\bbox{\mu}$'s components that correspond to the $H_{\kappa}$ involving $z_j$
and $z_k$. Because we chose to gather all the $z$-coordinates after all 
$(xy)$-pairs in (\ref{arrangement}), the Jacobi vector indices $j,k$ are
shifted relative to $\bbox{\mu}$'s components for the $z$-coordinates.
To simplify the notation, we use 
\begin{equation}
\iota\equiv N-1+\left\lfloor\frac{j+1}{2}\right\rfloor;\qquad 
\kappa\equiv N-1+\left\lfloor\frac{k+1}{2}\right\rfloor,
\end{equation}
so that $H_\iota$ and $m_\iota$ now refer to the operator and quantum number
pertaining to the coordinate $z_j$, with the same connection between 
$H_\kappa$, $m_\kappa$, and $z_k$.
Depending on the indices $j$ and $k$, i.e., depending on the positions of the
coordinates $z_j$ and $z_k$ in the sequence (\ref{arrangement}), $J_{z_jz_k}$ 
takes alternative forms in terms of ladder operators or $H_\kappa$.
The possible cases are:
\begin{enumerate}
\item $k$ even and $j=k-1$. According to (\ref{Hzz}), $J_{z_jz_k}$ 
coincides with $H_{N-1+k/2}\equiv H_{\kappa}$. The rotation matrix element is 
simply
\begin{equation}
%
%
\langle Y_{\Lambda',\bbox{\mu}',\bbox{\lambda}'}|\exp(i\phi\,J_{z_jz_k})|
Y_{\Lambda,\bbox{\mu},\bbox{\lambda}}\rangle=
%
%
\exp(im_{\kappa}\phi)\,
\delta_{\Lambda',\Lambda}\delta_{\bbox{\lambda}',\bbox{\lambda}}
\delta_{\bbox{\mu}',\bbox{\mu}}.
%
%
\end{equation}
\item $k$ even and $j<k-1$ even. We find 
\begin{equation}\label{eveneven}
J_{z_jz_k}=\frac{1}{4}\left(E_{\iota\kappa}^{+-}+E_{\iota\kappa}^{-+}+
E_{\iota\kappa}^{++}+E_{\iota\kappa}^{--}\right).
\end{equation}
Viewing again the sum of a ladder operator and its inverse (Hermitian
conjugate) as proportional to an analogue of a rotation about the $x$-axis,
we readily recognize the last expression as the sum of {\em two} $x$-type 
rotations (compare with (\ref{rotxy})'s central expression).
With the explicit formulae (\ref{updown}--\ref{upup}), 
it is straightforward to verify that the ladder operator 
$E_{\iota\kappa}^{+-}$ raising $m_\iota$ and lowering $m_\kappa$ commutes with
the ``raising-raising'' operator $E_{\iota\kappa}^{++}$. The two
$x$-type rotations are therefore independent of each other. As in the 
previous discussion of the $x$- and $y$-parts, an integer number of steps
$E_{\iota\kappa}^{+-}$ must connect $(m_\iota,m_\kappa)$ to 
$(m_\iota',m_\kappa')$, i.e., $m_\iota'=m_\iota+n_1,\; m_\kappa'=m_\kappa-n_1$.
However, the second rotation involves the same components of $\bbox{\mu}$, only
this time changing both in the {\em same} direction: $m_\iota'=m_\iota+n_2,\;
m_\kappa'=m_\kappa+n_2$. Compatibility of the two conditions requires 
$n_1=n_2=0$, thus
\begin{mathletters}
\begin{eqnarray}
m_{\bbox{\beta}_-(\iota\kappa)}&=&\frac{1}{2}(m_\iota-m_\kappa)=
m'_{\bbox{\beta}_-(\iota\kappa)} \\
m_{\bbox{\beta}_+(\iota\kappa)}& =& \frac{1}{2}(m_\iota+
m_\kappa)=m'_{\bbox{\beta}_+(\iota\kappa)},
\end{eqnarray}
\end{mathletters}
and the extra phase factors occurring in (\ref{xypart}) drop out in this
case:
\begin{equation}
%
%
\langle Y_{\Lambda',\bbox{\mu}',\bbox{\lambda}'}|\exp(i\phi\,J_{z_jz_k})|
Y_{\Lambda,\bbox{\mu},\bbox{\lambda}}\rangle=
d_{m_{\bbox{\beta}_-},m_{\bbox{\beta}_-}}^{(\lambda_{\bbox{\beta}_-})}
(\phi/2)
%
%
d_{m_{\bbox{\beta}_+},m_{\bbox{\beta}_+}}^{(\lambda_{\bbox{\beta}_+})}
(\phi/2)\, 
\delta_{\Lambda',\Lambda}\delta_{\bbox{\lambda}',\bbox{\lambda}}
\delta_{\bbox{\mu}',\bbox{\mu}}.\label{matee}
%
%
\end{equation}
The factor $1/2$ multiplying the rotation angle stems of course from the
factor $1/4$ in (\ref{eveneven}).
\item $k$ even and $j<k-1$ odd. In this case, 
\begin{equation}
J_{z_jz_k}=\frac{1}{4i}\left(E_{\iota\kappa}^{+-}-E_{\iota\kappa}^{-+}+
E_{\iota\kappa}^{++}-E_{\iota\kappa}^{--}\right),
\end{equation}
obviously the sum of two $y$-type rotations. The ``selection rules'' for 
$m_\iota$ and $m_\kappa$ are the same as in the previous case, and since
the extra phase factors distinguishing an $x$- from a $y$-type rotation 
cancel in (\ref{matee}), the {\em matrix element is identical to
the one in case~2}.
\item $k<N-1$ odd and $j$ even. $J_{z_jz_k}$ turns into
\begin{equation}
J_{z_jz_k}=\frac{1}{4i}\left(E_{\iota\kappa}^{+-}-E_{\iota\kappa}^{-+}-
E_{\iota\kappa}^{++}+E_{\iota\kappa}^{--}\right).
\end{equation}
Exactly the same considerations apply again, except that the {\em difference}
of two $y$-type rotations corresponds to a rotation through the angle $-\phi/2$
for the $E_{\iota\kappa}^{++}$ part.
\item $k<N-1$ odd and $j$ even, yielding
\begin{equation}
J_{z_jz_k}=\frac{1}{4}\left(E_{\iota\kappa}^{+-}+E_{\iota\kappa}^{-+}-
E_{\iota\kappa}^{++}-E_{\iota\kappa}^{--}\right),
\end{equation}
i.e., the difference of two $x$-type rotations. The matrix element
coincides with the one of case~4.
\item $k=N-1$ odd and $j$ even. $z_k$ does not occur in any of the $H_\kappa$, 
and the relevant ladder operators act on 
$m_\iota$ only:
\begin{equation}
J_{z_jz_k}=\frac{1}{2}\left(E_\iota^++E_\iota^-\right).
\end{equation}
This is an $x$-type rotation, exactly as in three dimensions. 
Projecting $\bbox{\mu}$ and $\bbox{\mu}'$ onto the appropriate
direction $\bbox{\gamma}(\iota)$: 
$m_{\bbox{\gamma}(\iota)}=m_\iota,\; m_{\bbox{\gamma}(\iota)}'=
m_\iota'=m_{\bbox{\gamma}(\iota)}+\nu$, 
we obtain for the matrix element
\begin{equation}
%
%
\langle Y_{\Lambda',\bbox{\mu}',\bbox{\lambda}'}|\exp(i\phi\,J_{z_jz_k})|
Y_{\Lambda,\bbox{\mu},\bbox{\lambda}}\rangle=
d_{m_{\bbox{\gamma}}',m^{}_{\bbox{\gamma}}}^{(\lambda_{\bbox{\gamma}})}
(\phi)
%
%
e^{i(m_{\bbox{\gamma}}'-m^{}_{\bbox{\gamma}})\pi/2}
\delta_{\Lambda',\Lambda}\delta_{\bbox{\lambda}',\bbox{\lambda}}
\delta_{\bbox{\mu}',\bbox{\mu}+\nu\bbox{\gamma}}.
%
%
\end{equation}
\item $k=N-1$ odd and $j$ odd. 
\begin{equation}
J_{z_jz_k}=\frac{1}{2i}\left(E_\iota^+-E_\iota^-\right),
\end{equation}
a $y$-type rotation. Its matrix element differs from the one in case~6 only
by the absence of the phase factor $\exp(i\nu\pi/2)$.
\end{enumerate}
Thus, the $z$-part of the rotation~(\ref{finiterot}) takes essentially three
different forms: (i) a phase factor $\exp(im_\kappa\phi)$, diagonal in the
$\bbox{\mu}$, if $J_{z_jz_k}$ coincides with $H_\kappa$; (ii) an $x$-
or $y$-type rotation through the angle $\phi$, changing $m_\iota$ into
$m_\iota'=m_\iota+\nu$, as in three dimensions, if $z_k$ is the unpaired
coordinate $z_{N-1}$ of an odd-dimensional space; (iii) the product of two
rotations
through the angles $\phi/2$ and $(-1)^k\phi/2$, respectively, but diagonal in
$\bbox{\mu}$, if $z_j$ and $z_k$ belong to different $H_\iota$ and $H_\kappa$.
\subsection{Harmonics suitable for rotation}
\label{harmconst}
In the previous subsection, we have completely determined the matrix
elements for the rotations of our interest when applied to hyperspherical
harmonics, before actually specifying these functions explicitly. This was
possible because we expressed the matrix elements in terms of the
labels (``quantum numbers'') $(\Lambda,\bbox{\mu},\bbox{\lambda})$ identifying 
the harmonics, rather than through integrals in coordinate space.
So far, the harmonics are functions of the generic Cartesian coordinates
$(x_j,y_j,z_j)$. Upon rotation from one Jacobi tree to another, these 
coordinates transform into another set $(x_j',y_j',z_j'),\;j=1,\ldots,N-1$, but
the harmonics, when expressed in terms of the new coordinates, retain their
functional form. In this sense, these hyperspherical harmonics are
frame-independent.

Direct solution of the angular Laplacian's eigenvalue problem by separation
of the $(3N-4)$ coordinates in the second-order differential equation leads 
to hyperspherical harmonics represented by standard spherical harmonics and 
Jacobi or Gegenbauer polynomials, depending on the choice of hyperspherical
coordinates \cite{Smirnov,Cavagnero,Avery}. However, these harmonics are not
simultaneous eigenfunctions of all the $H_j$. Consequently, they are not
suitable for our purpose, because rotating the harmonics 
requires knowledge of the ladder operators' effects on these functions, at
least for the base set of ladder operators $E_{\pm\bbox{\epsilon}_j}$, 
$j=1,\ldots,\ell$, which in turn requires uniquely specifying the harmonics
with labels $(\Lambda,\bbox{\mu},\bbox{\lambda})$. We now construct complete
sets of functions defined 
exclusively by their behavior under the
action of the first-order differential operators $H_j$ and 
$E_{\pm\bbox{\epsilon}_j}$, $j=1,\ldots,\ell$. 
The resulting functions are ``hyperspherical harmonics'', too, because 
they satisfy the generalized Laplacian's symmetry under rotations. 

The hyperspherical description separates the ``hyperradial'' momentum from
the generalized angular momentum. 
Each of the Cartesian coordinates $x_j,y_j,z_j$, ($j=1,\ldots,N-1$), is 
proportional to the ``hyperradius'' $R$. The angular Laplacian's eigenvalue 
parameter $\Lambda$ determines the harmonic's {\em degree} by setting the
radial scale as $R^\Lambda$. Neither the $H_j$ nor the ladder operators 
$E_{\pm\bbox{\epsilon}_j}$ affect this radial factor, as is to be expected 
of angular momentum-like operators. With $(3N-4)$ angular coordinates, a 
complete specification of the harmonics requires $(3N-5)$ labels  in addition 
to $\Lambda$. The vector $\bbox{\mu}$ provides $\ell=\lfloor (3N-3)/2\rfloor$ 
of them in the form of the eigenvalues $m_j$ of all the $H_j$. The remaining 
labels are taken from the $\ell$-component vector $\bbox{\lambda}$ consisting 
of the ``string lengths'' $\lambda_{\bbox{\epsilon}_j}$ for the ladder 
operators $E_{\pm\bbox{\epsilon}_j}$. The first
$(\ell-2)$ components of $\bbox{\lambda}$ suffice to reach a total of $(3N-4)$
labels if $N$ is odd, as do the 
first $(\ell-1)$ components for $N$ even. 

It remains to determine the function $F$ introduced in (\ref{efHi}).
The requirement $H_j\,F\equiv 0$ for all $j$ implies that $F$ depends only 
on $(u_j^2+v_j^2)$ (and possibly on the single unpaired coordinate $w$ in case
$(3N-3)$ is odd). Furthermore, if $\sum_j|m_j|=\Lambda$ the product
$\prod_j(u_j+i\,v_j)^{m_j}$ already accounts for the radial factor $R^\Lambda$,
i.e., $F=\mathrm{const.}$ in this case. This circumstance suggests constructing
complete sets of degenerate harmonics with the same $\Lambda$ as follows:
We set $m_1$ to its maximum value and all other $m_j=0$, i.e., 
$\bbox{\mu}=(\Lambda,0,\ldots,0)$ \cite{irrep}. 
For $j=1,\ldots,\ell$, the modulus of the projection
$m_{\bbox{\epsilon}_j}=\bbox{\mu}\cdot\bbox{\epsilon}_j/
(\bbox{\epsilon}_j\cdot\bbox{\epsilon}_j)$
sets the lower limit for the corresponding string length 
$\lambda_{\bbox{\epsilon}_j}$. However, $\lambda_{\bbox{\epsilon}_j}$  
cannot be larger than $m_{\bbox{\epsilon}_j}$ either, for applying any of 
the {\em raising} operators $E_{+\bbox{\epsilon}_j}$ necessarily results in 
a vector $\bbox{\mu}'=\bbox{\mu}+\bbox{\epsilon}_j$ having 
$\sum_j|m_j'|>\Lambda$, i.e., in a harmonic not belonging to the same set of 
degenerate functions. With $\bbox{\lambda}$ completely specified, the first 
harmonic reads 
%
%
\begin{equation}
\label{firstY}
Y_{\Lambda,\bbox{\mu},\bbox{\lambda}}
%
%
=c_\Lambda\,(x_1+i\,y_1)^\Lambda;\qquad \bbox{\lambda}=
(m_{\bbox{\epsilon}_1},0,\ldots,0);\qquad\bbox{\mu}=(\Lambda,0,\ldots,0),
%
%
\end{equation}
where $c_\Lambda$ denotes a normalization constant. Starting from this first
function, we generate harmonics by applying all the 
lowering operators $E_{-\bbox{\epsilon}_j}$ first to (\ref{firstY}), then
applying the $E_{-\bbox{\epsilon}_j}$ to the harmonics so obtained,
and so on in a {\em recursive procedure}: 
\begin{equation}
Y_{\Lambda,\bbox{\mu}-\bbox{\epsilon}_j,\bbox{\lambda}'}=\frac{1}{\sqrt{(
\lambda_{\bbox{\epsilon}_j}+m_{\bbox{\epsilon}_j})(\lambda_{\bbox{
\epsilon}_j}-
m_{\bbox{\epsilon}_j}+1)}}\, E_{-\bbox{\epsilon}_j}Y_{\Lambda,
\bbox{\mu},\bbox{\lambda}}
\end{equation}
for all $\bbox{\epsilon}_j$-strings that have not yet terminated, i.e., 
the strings having $m_{\bbox{\epsilon}_j}>-\lambda_{\bbox{\epsilon}_j}$. 
The recursive nature of 
the process suggests gathering together the harmonics in ``levels'', with   
the level of a harmonic indicating the number of lowering operator steps 
separating it from the first harmonic (\ref{firstY}). At level 0 with  the
harmonic (\ref{firstY}), the 
$\bbox{\epsilon}_1$-string is the only string with non-zero length, and we 
can only 
generate one harmonic of level 1. But for $\ell>1$, this level-1 harmonic 
will already have several non-vanishing $m'_{\bbox{\epsilon}_j}$
and corresponding $\lambda'_{\bbox{\epsilon}_j}$,
giving rise to additional $\bbox{\epsilon}_j$-strings starting from this 
level.
Repeating this procedure level by level, we work our way down until finally
reaching the last harmonic with $\bbox{\mu}=(-\Lambda,0,\ldots,0)$. The 
procedure stops automatically, because applying any of the lowering operators
to this last harmonic maps it to zero (exactly as $l_-Y_{l,-l}\equiv 0$ in
three dimensions). 

In $d$ dimensions, the total number of different
hyperspherical harmonics with the same ``grand angular momentum'' $\Lambda$
is \cite[p. 265]{Rau}
\begin{equation}
{\rm dim}(\Lambda;\,d)=
\frac{2\Lambda+d-2}{\Lambda+d-2}{\Lambda+d-2 \choose d-2}.
\label{dimlambda}
\end{equation}
For a given $\Lambda$, the procedure outlined above generates exactly
${\rm dim}(\Lambda,\,d)$ independent functions, i.e., a complete set of 
hyperspherical harmonics. However, some of these harmonics still require  
modifying for our purpose, as the following observations illustrate. 

Note first that with our choice for $E_{\bbox{\epsilon}_j}$, 
the first $(\ell-1)$ ladder operators always change two
$m_j$ in opposite directions, thus leaving $\sum_j|m_j|=\Lambda$ invariant
when starting from the first harmonic. For all these harmonics, the function
$F$ reduces to a constant. 
At some stage during the above construction, however, an
$\bbox{\epsilon}_{\ell}$-string will appear along which the sum 
$\sum_j|m_j|$ no longer remains constant.
Since the ladder operators do not change the overall radial factor $R^\Lambda$,
the function $F$ accounts for any ``missing powers'' in 
$\prod_j(u_j+iv_j)^{m_j}$  
for harmonics having $\bbox{\mu}$ with $\sum_j|m_j|<\Lambda$. For instance,
in three dimensions the lowering operator $l_-$ ($E_1^-$ in (\ref{Lpm})'s
notation) removes one power of $(x+iy)$ from the ``highest'' 
harmonic $r^lY_{ll}\sim (x+iy)^l=r^l\sin^l\theta\exp(il\phi)$
while simultaneously adding a factor $F=z=r\cos\theta$ to the next harmonic 
$r^lY_{l,l-1}\sim
z\,(x+iy)^{l-1}=r^l\cos\theta\sin^{l-1}\theta\exp(i[l-1]\phi)$.
In the same way, repeated application of $l_-$ generates the higher-degree 
Legendre polynomials in $\cos\theta$ making up $F$ in this case.

So far, $\Lambda$ and the vector $\bbox{\mu}$ provide enough information to
uniquely specify the hyperspherical harmonics; we need their additional label
$\bbox{\lambda}$ only for the purpose of determining transformation matrix
elements, not to distinguish the harmonics from one another. 
For these harmonics, the construction described above provides the appropriate 
function $F$. 

In a higher-dimensional setting, however, the $E_{\bbox{\epsilon}_j}$ do 
not all commute with each other.
It is thus possible to arrive at {\em different} functions with the {\em same}
$\bbox{\mu}$ along different ladder operator sequences, starting from the
first harmonic (\ref{firstY}). This is exactly the situation of ``degenerate
eigenvalues'': in this case, $\bbox{\mu}$, the set of eigenvalues of the $H_j$,
has multiplicity higher than one. Although our construction generates the
appropriate number of independent functions, thus providing a basis system
for the higher-dimensional eigenspace associated with $\bbox{\mu}$, we now
need additional labels---to be taken from $\bbox{\lambda}$---to distinguish
the different eigenfunctions with degenerate eigenvalue $\bbox{\mu}$.
Suppose, therefore, the vector $\bbox{\mu}$ occurs with multiplicity $\nu>1$.
Our procedure generates $\nu$ independent functions
$\Phi_\rho,\;\rho=1,\ldots,\nu$, all having the same
$\prod_j(u_j+iv_j)^{m_j}$. These functions differ only in their $F$.
Harmonics suitable for the calculation of rotation matrix
elements are expressed as linear combinations 
\begin{equation}
\tilde{\Phi}_\sigma=\sum_{\rho=1}^\nu a_{\sigma\rho}\Phi_\rho,
\end{equation}
and the requirement 
\begin{equation}
\left(E_{+\bbox{\epsilon}_j}\right)^{\lambda_{\bbox{\epsilon}_j}-
    m_{\bbox{\epsilon}_j}+1}
\,\tilde{\Phi}_\sigma=0,\qquad\text{for }j=1,\ldots,j_{\rm max}
\end{equation}
determines the coefficients $a_{\sigma\rho}$ for $\sigma=1,\ldots,\nu$
(and thus the functions $F$). In this way, appropriate sets of parameters
$(\lambda_{\bbox{\epsilon}_1},\ldots,\lambda_{\bbox{\epsilon}_{j_{\rm 
      max}}})$ 
serve to distinguish the harmonics by enforcing specific lengths for the 
different $\bbox{\epsilon}_j$-strings.
As noted previously, the number of additional labels required is
$j_{\rm max}=\ell-2$ (or $\ell-1$) for even (or odd)-dimensional spaces,
respectively. With this modification, even the harmonics corresponding to
degenerate vectors $\bbox{\mu}$ show the desired behavior under rotations
in $d$ dimensions, and their transformation matrix elements can be deduced
from their ``quantum numbers'' $(\Lambda,\bbox{\mu},\bbox{\lambda})$ directly.

Finally, a remark concerning the degeneracy of $\bbox{\mu}$ seems in order. In 
three dimensions, it is impossible to arrive at the same $\bbox{\mu}\equiv m$
along different strings, because there is {\em only one pair of ladder
operators}. Nevertheless, even in this case we note that the $m$-components
arising for a given $l$ ($\equiv\Lambda$) occur again for $l'>l$. Due to the
mostly pairwise change of $m_j$ in higher-dimensional spaces, we expect that
a vector $\bbox{\mu}$ occurring for a given $\Lambda$ will appear again as a
label for harmonics with $\Lambda'=\Lambda+2,\Lambda+4,\ldots$. For $d>4$ 
there are non-commuting ladder operators. Because the
number of different pathways leading from the highest harmonic to a
specific $\bbox{\mu}$
increases with the length of this path,
$\bbox{\mu}$'s multiplicity increases with $\Lambda$. 
Furthermore, since different functions $F$ imply contributions
from different $(u_j^2+v_j^2)$-terms, higher multiplicity---even in spaces
with more than three dimensions---can only arise for $\Lambda-\sum_j|m_j|\ge2$.
A detailed analysis of the recursion in $\Lambda$ with $\bbox{\mu}$ held 
fixed confirms these expectations, yielding for the multiplicity of
$\bbox{\mu}$ 
\begin{eqnarray} 
{\rm mult}(\bbox{\mu};\;\Lambda,\,d)& =& { p+q\choose q}
\label{multipl} \\ \text{with} \qquad
  p&=&\left\lfloor\frac{\Lambda-\sum_j|m_j|}{2}\right\rfloor;
\quad q= \left\lfloor \frac{d-3}{2}\right\rfloor, \nonumber
\end{eqnarray}
a useful result to test the implementation of the
(recursive) procedure.
\section{Example: Two-electron system}
\label{example}
For the purpose of illustration, we apply the method outlined in the
preceding sections to a Coulombic three-body system. Specifically, we
consider a two-electron atom or ion with atomic number $Z$, i.e., a system 
with only one heavy 
particle. The case of diatomic molecules (two-center Coulomb system) requires
additional modifications of the hyperspherical approach\cite{Tolstikhin}.
\subsection{Coulomb interactions among three particles}
After elimination of the CM motion, a three-particle system requires two 
Jacobi vectors for its description.
We use three different Jacobi trees, $\bbox{\xi}_i$, $\bbox{\eta}_i$, 
$\bbox{\zeta}_i$, $(i=1,2)$.
With $\bbox{r}_1$, $\bbox{r}_2$, and $\bbox{r}_N$ denoting the positions of the
two electrons and the nucleus, respectively, the relevant mass-weighted
Jacobi vectors read
\begin{mathletters}
\begin{eqnarray}
\bbox{\xi}_1 & = & \sqrt{\frac{M_e}{2}}(\bbox{r}_1-\bbox{r}_2), \\
\bbox{\xi}_2 & = & 
\sqrt{\frac{2M_eM_N}{M_N+2M_e}}\left(\frac{\bbox{r}_1
+\bbox{r}_2}{2}-\bbox{r}_N\right); 
\end{eqnarray}
\end{mathletters}
\begin{eqnarray}
\bbox{\eta}_1 & = & \sqrt{\frac{M_NM_e}{M_N+M_e}}(\bbox{r}_N-\bbox{r}_1)
%
%
 = \cos\beta\;\bbox{\xi}_1+\sin\beta\;\bbox{\xi}_2; \label{eta1} \\
\bbox{\zeta}_1 & = & \sqrt{\frac{M_NM_e}{M_N+M_e}}(\bbox{r}_2-\bbox{r}_N)
%
%
 = \cos\gamma\;\bbox{\xi}_1+\sin\gamma\;\bbox{\xi}_2; \label{zeta1}
\end{eqnarray} 
where
\begin{mathletters}
\begin{eqnarray}
\cos\beta & = & -\sqrt{\frac{M_N}{2(M_N+M_e)}},\\
\sin\beta & = & -\sqrt{\frac{M_N+2M_e}{2(M_N+M_e)}},
\end{eqnarray}
\end{mathletters}
and $\gamma=-\beta$. $M_N$ and $M_e$ denote the nuclear and electron mass,
respectively.
The vectors $\bbox{\eta}_2$ and $\bbox{\zeta}_2$ will not
be needed, since the Coulomb interaction among the three particles takes the 
form
\begin{equation}
V_C=\sqrt{\frac{M_e}{2}}\,
  \frac{1}{|\bbox{\xi}_1|}-\sqrt{\frac{M_NM_e}{M_N+M_e}}\left(
  \frac{Z}{|\bbox{\eta}_1|}+\frac{Z}{|\bbox{\zeta}_1|}\right).
\end{equation}
At this point, the familiar approach using the transformation of the 
{\em coordinates}
would exploit (\ref{eta1}--\ref{zeta1}) in a multipole expansion
of the electron-nucleus interactions in terms of the coordinates $\bbox{\xi}_1$
and $\bbox{\xi}_2$. In our method, however, we apply the (kinematic) rotations
(\ref{eta1}--\ref{zeta1}) to the {\em wave functions} instead. This 
amounts to calculating the three interaction matrix elements in three 
different coordinate systems. The immediately obvious advantage of this 
method is that each of the three terms takes exactly the same form. Each of 
the integrals reduces
to the same
radial integral as in the textbook example of hydrogen. Higher-order multipoles
and nested integrations over powers of the coordinates do not occur.
\subsection{Choice of $H_j$ and ladder operators}
The kinetic energy has rotational symmetry in six dimensions. 
Partitioning the six coordinates into pairs defines three  
non-intersecting planes, and thus $\ell=3$ mutually commuting rotations.
We choose the corresponding first-order operators $H_j$
as $J_{x_1y_1},\;J_{x_2y_2}$, and $J_{z_1z_2}$. Here 
$\{x_1,y_1,z_1,x_2,y_2,z_2\}$ denote the Cartesian components
of the mass-weighted Jacobi vectors. Accordingly, the 
harmonics---simultaneous eigenfunctions of the three $H_j$ with eigenvalues
$\bbox{\mu}=(m_1,m_2,m_3)$---take the form 
\begin{equation}
F(x_1^2+y_1^2,\, x_2^2+y_2^2,\, z_1^2+z_2^2)\;\prod_{j=1}^2
(x_j+i{\rm sign}(m_j)\,y_j)^{|m_j|} \;
(z_1+i{\rm sign}(m_3)\,z_2)^{|m_3|}, \label{so6harm}
\end{equation}
where we have rewritten (\ref{efHi}) so as to yield non-negative powers of
the hyperradius for either sign of the $m_j$. The appropriate base set 
of three ladder operator pairs $E_{\pm\bbox{\epsilon}_j}$ is then specified 
by the vector labels 
$\bbox{\epsilon}_1=(1,-1,0)$, $\bbox{\epsilon}_2=(0,1,-1)$, and 
$\bbox{\epsilon}_3=(0,1,1)$, corresponding to 
$E_{12}^{\pm\mp},\;E_{23}^{\pm\mp}$, and $E_{23}^{\pm\pm}$ in 
(\ref{updown}--\ref{upup}).
\subsection{Labeling the basis functions}
Five angles specify each point on a sphere of fixed hyperradius $R$ in
six-dimensional space. Consequently, our hyperspherical harmonics require
five labels---related to the numbers of nodes in the five angular
variables---for their identification. Besides $\Lambda$ and $\bbox{\mu}=
(m_1,m_2,m_3)$ we need one more label, $\lambda_{\bbox{\epsilon}_1}$, 
the string length along
the ladder operator sequences spanned by $E_{\pm\bbox{\epsilon}_1}$.

The set of harmonics so defined differs  from the more familiar
set labeled by quantum numbers $(l_1,m_1,l_2,m_2,n_\alpha)$ 
\cite{Cavagnero,Rau}.
However, the latter set of harmonics does not make use of the third eigenvalue
$m_3$. Since rotating the harmonics is accomplished by acting on them with
ladder operators which in turn act on {\em all} the $m_j$, the conventional
harmonics specified by $(l_1,m_1,l_2,m_2,n_\alpha)$ are not suited for our
purpose. They may, of course, still serve as a basis set in an application,
being then expanded into the new set labeled by 
$(\Lambda,\bbox{\mu},\lambda_{\bbox{\epsilon}_1})$
prior to the actual rotation. A straightforward transformation links the two
basis sets.
\subsection{Explicit construction of harmonics with degenerate $\bbox{\mu}$}
As an example, we derive the expressions for some harmonics that are not
completely characterized by $\Lambda$ and $\bbox{\mu}$. Consider for instance
harmonics with $\Lambda=4$, $\bbox{\mu}=(1,0,1)$. According to 
(\ref{multipl}), there are two harmonics with these labels, to be
distinguished by the additional parameter $\lambda_{\bbox{\epsilon}_1}$.
With the particular vector $\bbox{\mu}$ of this example, we obtain 
$m_{\bbox{\epsilon}_1}=\frac12$, thus setting the lower limit for 
$\lambda_{\bbox{\epsilon}_1}$. The two essentially different ways of arriving
at the set of labels $(1,0,1)$ from harmonics of the next-lower level are
given by an $E_{-\bbox{\epsilon}_1}$-step from $\bbox{\mu}'=(2,-1,1)$, and
by an $E_{-\bbox{\epsilon}_3}$-step from $\bbox{\mu}''=(1,1,2)$. Both of these
labels have multiplicity 1 since $\sum_j|m_j|=\Lambda$; the function $F(\ldots)$
in the corresponding harmonics reduces to a normalization constant. Up to
these normalizing factors, $c_1$ and $c_2$, (\ref{so6harm}) gives these 
harmonics as 
\begin{mathletters}
\begin{eqnarray} 
Y_{4,(2,-1,1),\frac32} & = & c_1\,(x_1+iy_1)^2\,(x_2-iy_2)
(z_1+iz_2), \\
Y_{4,(1,1,2),1} & = & c_2\,(x_1+iy_1)\,(x_2+iy_2)
(z_1+iz_2)^2.
\end{eqnarray}
\end{mathletters}
Applying the appropriate ladder operators to these harmonics
provides a basis for the two-dimensional eigenspace of degenerate 
$\bbox{\mu}=(1,0,1)$:
\begin{mathletters}
\begin{eqnarray}
\Phi_1 & = & E_{-\bbox{\epsilon}_1} Y_{4,(2,-1,1),\frac32}
\nonumber \\ & = & c_3\,(x_1+iy_1)(z_1+iz_2)\;([x_1^2+y_1^2]-2[x_2^2+y_2^2])
\\ 
\Phi_2 & = & E_{-\bbox{\epsilon}_3} Y_{4,(1,1,2),1}
\nonumber \\ & = & c_4\,(x_1+iy_1)(z_1+iz_2)\;(2[x_2^2+y_2^2]-[z_1^2+z_1^2]),
\end{eqnarray}
\end{mathletters}
with constant factors $c_3$, $c_4$ to be ultimately absorbed into the 
normalization.
Because $\Phi_1$ is obtained from $Y_{4,(2,-1,1),\frac32}$ by applying 
$E_{-\bbox{\epsilon}_1}$, it behaves under the relevant rotations 
exactly as is required for $Y_{4,(1,0,1),\frac32}$. Apart from the 
normalization constant $c_3$, we thus find 
\begin{equation}
Y_{4,(1,0,1),\frac32} = 
c_3\,(x_1+iy_1)(z_1+iz_2)\;([x_1^2+y_1^2]-2[x_2^2+y_2^2]).
\end{equation}
However, the second harmonic with the same $\bbox{\mu}=(1,0,1)$ does not
simply coincide with $\Phi_2$, because the latter has the same string-length
$\lambda_{\bbox{\epsilon}_1}=\frac32$ as $\Phi_1$: One easily verifies
that 
\begin{equation}
E_{\bbox{\epsilon}_1}\Phi_2\neq 0.
\end{equation}
We can, however, construct a harmonic $Y_{4,(1,0,1),\frac12}$ as a linear
combination of $\Phi_1$ and $\Phi_2$ by requiring
\begin{equation}
E_{\bbox{\epsilon}_1} (a\Phi_1+b\Phi_2) \equiv 0,
\end{equation}
yielding the condition $3ac_3-2bc_4=0$ and thus
\begin{equation}
Y_{4,(1,0,1),\frac12}= c_5 (x_1+iy_1)(z_1+iz_2)\;(2[x_1^2+y_1^2]+2[x_2^2+y_2^2]
-3[z_1^2+z_2^2]).
\end{equation}
The two harmonics with ``degenerate'' $\bbox{\mu}=(1,0,1)$ are now 
distinguished by their respective $\lambda_{\bbox{\epsilon}_1}$.
\subsection{Rotating the harmonics}
Eq.~(\ref{eta1}) describes the {\em coordinate} transformation between the 
Jacobi trees $(\bbox{\xi}_1,\bbox{\xi}_2)$ and $(\bbox{\eta}_1,\bbox{\eta}_2)$.
Accordingly, the hyperspherical harmonics transform as
\begin{equation}
|\Lambda,\bbox{\mu},\lambda_{\bbox{\epsilon}_1}\rangle_{\bbox{\xi}} 
= \sum_{\bbox{\mu}'} D_{\bbox{\mu}',\bbox{\mu}}^{(\lambda_{\bbox{\epsilon}_1})}
(\beta) \;
|\Lambda,\bbox{\mu}',\lambda_{\bbox{\epsilon}_1}\rangle_{\bbox{\eta}}
\end{equation}
where the subscripts on the ket vectors indicate the respective Jacobi tree.
The transformation matrix elements are given by
\begin{eqnarray}
D_{\bbox{\mu}',\bbox{\mu}}^{(\lambda_{\bbox{\epsilon}_1})}(\beta) & = &
\langle\Lambda,\bbox{\mu}',\lambda_{\bbox{\epsilon}_1}| 
\exp(i\beta[J_{x_1x_2}+
J_{y_1y_2}+J_{z_1z_2}]) |\Lambda,\bbox{\mu},\lambda_{\bbox{\epsilon}_1}\rangle 
\nonumber\\ & = &
\langle\Lambda,\bbox{\mu}',\lambda_{\bbox{\epsilon}_1}| 
\exp(i\beta[E_{\bbox{\epsilon}_1}+
E_{-\bbox{\epsilon}_1}+H_3]) |\Lambda,\bbox{\mu},\lambda_{\bbox{\epsilon}_1}
\rangle ,
\end{eqnarray}
according to Subsec.~\ref{rotharms}. Following the analysis given there, we
find 
\begin{equation}
D_{\bbox{\mu}',\bbox{\mu}}^{(\lambda_{\bbox{\epsilon}_1})}(\beta) = 
e^{i[m'-m]\pi/2}d_{m',m}^{(\lambda_{\bbox{\epsilon}_1})}(2\beta)\;
e^{im_3\beta}\;\delta_{\bbox{\mu}',\bbox{\mu}+n\bbox{\epsilon}_1},
\end{equation}
with 
\begin{mathletters}
\begin{eqnarray}
m & = & \frac12 (m_1-m_2) \\
m' & = & \frac12 (m_1'-m_2') \\
n & = & m_1'-m_1 = -m_2'+m_2 
\end{eqnarray}
\end{mathletters}
in terms of $\bbox{\mu}$-components. Replacing the angle $\beta$ with 
$\gamma=-\beta$ yields the expressions for the transformation to Jacobi tree
$\bbox{\zeta}$.
\subsection{Symmetry properties of basis functions}
Due to the very high degeneracy of harmonics with the same grand angular
momentum $\Lambda$ as expressed in (\ref{dimlambda}), it is important to
exploit various symmetries of the functions in order to reduce the size of the
hyperspherical basis. Reflection through the origin of the coordinate system
transforms all six coordinates into their negatives. According to
(\ref{so6harm}), the harmonic 
$|\Lambda,\bbox{\mu},\lambda_{\bbox{\epsilon}_1}\rangle$ picks up a
factor $(-1)^{|m_1|+|m_2|+|m_3|}$ under this operation. Since all the ladder
operators change two of the $m_j$ at a time, the sum in the exponent has
the same parity as $\Lambda$. This first observation thus restricts the
basis set by allowing only even $\Lambda$ for even-parity states and
odd $\Lambda$ for odd-parity states.

Consider next the harmonics' symmetry under interchange of identical particles
(i.e., the two electrons). This interchange is most easily described in the 
Jacobi tree $\{\bbox{\xi}_1,\bbox{\xi}_2\}$:
\begin{equation}
\bbox{\xi}_1\leftrightarrow -\bbox{\xi}_1;\qquad  
\bbox{\xi}_2\leftrightarrow \bbox{\xi}_2.
\end{equation}
Accordingly, the harmonic $|\Lambda,m_1,m_2,m_3,\lambda_{\bbox{\epsilon}_1}
\rangle$ 
turns into the
harmonic $|\Lambda,m_1,m_2,-m_3,\lambda_{\bbox{\epsilon}_1}\rangle$ under 
interchange of the electrons
while acquiring a factor $(-1)^{m_1+m_3+S}$ (where $S$ denotes the total
spin of the two electrons). Antisymmetrized basis functions may thus be
labeled by non-negative $m_3$ only, and $m_3=0$ is possible only for harmonics
with $(m_1+S)$ {\em even}.

Furthermore, when choosing the first-order operators $H_j$ we have arranged
the six coordinates in a way that allows us to identify $m_1$ and $m_2$ with
the $z$-projections of the three-dimensional relative angular momenta
$\bbox{l}_1$ and
$\bbox{l}_2$. Therefore, the sum $m_1+m_2=m_{\rm tot}$ is the $z$-component of
the coupled (total) orbital angular momentum $L$.
The fact that the Coulomb interaction does not couple states with
different $m_{\rm tot}$ reduces the size of a basis consisting of
antisymmetrized harmonics accordingly. This point is particularly
interesting because the system's invariant $m_{\rm tot}$ restricts our basis,
even though the corresponding total orbital angular momentum $L$ is {\em not
defined} in this basis. The reason for this seemingly surprising fact lies
in our use of first-order operators only; hence $L_z$
can be identified, but not the second order operators $\bbox{l}_j^2$ or $L^2$.
The absence of the invariant three-dimensional $L$ is the main trade-off we 
have to accept when treating all transformations as rotations in a genuinely 
six-dimensional space \cite[especially Sec.~10.2., p.~267ff]{Rau}. 
\subsection{Calculation of matrix elements}
While Cartesian coordinates prove most appropriate for manipulating the
hyperspherical harmonics using ladder operators, hyperspherical coordinates
lend themselves for the calculation of matrix elements. Specifically,
the familiar representation of Cartesian coordinates
\begin{equation}
\begin{array}{rcl}
x_1 & = & R\cos\alpha\sin\theta_1\cos\varphi_1, \\
y_1 & = & R\cos\alpha\sin\theta_1\sin\varphi_1, \\
z_1 & = & R\cos\alpha\cos\theta_1,
\end{array}
\qquad\qquad
\begin{array}{rcl}
x_2 & = & R\sin\alpha\sin\theta_2\cos\varphi_2, \\
y_2 & = & R\sin\alpha\sin\theta_2\sin\varphi_2, \\
z_2 & = & R\sin\alpha\cos\theta_2, 
\end{array}
\label{HScoords}
\end{equation}
transforms (\ref{so6harm}) into
\begin{eqnarray}
& & R^\Lambda F\;\sin^{|m_1|}\theta_1 e^{im_1\varphi_1}
\sin^{|m_2|}\theta_2 e^{im_2\varphi_2}
%
%
(\cos\alpha\cos\theta_1+i{\rm sign}(m_3)\sin\alpha\cos\theta_2)^{|m_3|},
\end{eqnarray}
where $F$ is a function of 
$(\cos^2\alpha\sin^2\theta_1)$, $(\sin^2\alpha\sin^2\theta_2)$, and 
$(\cos^2\alpha\cos^2\theta_1+\sin^2\alpha\cos^2\theta_2)$. 
Note that this form remains the same, regardless of whether the Cartesian
components $(x,y,z)$ refer to the Jacobi vectors in tree $\bbox{\xi}$, in
$\bbox{\eta}$, or in $\bbox{\zeta}$. The relevant interaction operator 
entering into the matrix element is always $1/(R\cos\alpha)$, for each of the
three pairwise Coulomb interactions, with the angle $\alpha$ referring to 
a different coordinate system in each case. The $\theta$- and 
$\alpha$-integrals arising in the calculation are related to the Euler
Beta function \cite{Abramowitz}, namely,
\begin{mathletters}
\begin{eqnarray}
\int_0^{2\pi}d\phi\;e^{i(m'-m)\varphi} & = & 2\pi\delta_{m'm} \\
\int_0^{\pi}d\theta\;\sin^p\theta\;\cos^q\theta & = & \left(1+(-1)^q\right)
%
%
\frac{(p-1)!!(q-1)!!}{(p+q)!!}c_{pq} \\
\int_0^{\pi/2}d\alpha\;\sin^p\alpha\;\cos^q\alpha & = & 
\frac{(p-1)!!(q-1)!!}{(p+q)!!}c_{pq},\label{alphaint}
\end{eqnarray}
\end{mathletters}
where 
\begin{equation}
c_{pq}=\left\{\begin{array}{l@{\qquad}l} \frac{\pi}{2} & p, q {\rm \ both\  
even} \\
1 & {\rm otherwise}. \end{array}\right.
\end{equation}
A multipole expansion would have forced us to split the last
integral into two parts, with different integrands depending on whether
$|\cos\beta\bbox{\xi}_1|$ is greater or smaller than $|\sin\beta\bbox{\xi}_2|$.
The resulting integral could only be expressed as {\rm a sum} of factorial
quotients, rather than {\em a single term}, as in (\ref{alphaint}).

Gathering together all the pieces developed in this section, we find for
the angular part of the matrix elements 
\begin{eqnarray}
\langle\Lambda',\bbox{\mu}',\lambda_{\bbox{\epsilon}_1}'|V_C|
\Lambda,\bbox{\mu},\lambda_{\bbox{\epsilon}_1}\rangle & = & 
\langle\Lambda',\bbox{\mu}',\lambda_{\bbox{\epsilon}_1}'|
\frac{1}{r_{12}}-\frac{Z}{r_{1N}}-\frac{Z}{r_{2N}}
|\Lambda,\bbox{\mu},\lambda_{\bbox{\epsilon}_1}\rangle \\
& = & \frac{1}{R}\sqrt{\frac{M_e}{2}}
\langle\Lambda',\bbox{\mu}',\lambda_{\bbox{\epsilon}_1}'|\frac{1}{\cos
\alpha}|\Lambda,\bbox{\mu},\lambda_{\bbox{\epsilon}_1}\rangle - 
\frac{Z}{R} \sqrt{\frac{M_NM_e}{M_N+M_e}} \times\nonumber \\ & & \times
\sum_{\bbox{\mu}_1,\bbox{\mu}_2} \left\{
\left(D_{\bbox{\mu}_1,\bbox{\mu}'}^{(\lambda_{\bbox{\epsilon}_1}')}(\beta)
\right)^{\dagger}\; 
D_{\bbox{\mu}_2,\bbox{\mu}}^{(\lambda_{\bbox{\epsilon}_1})}(\beta)\;
+ \left(D_{\bbox{\mu}_1,\bbox{\mu}'}^{(\lambda_{\bbox{\epsilon}_1}')}(\gamma)
\right)^{\dagger}\;
D_{\bbox{\mu}_2,\bbox{\mu}}^{(\lambda_{\bbox{\epsilon}_1})}(\gamma)\;
\right\}\times\nonumber \\ & & \times
\langle\Lambda',\bbox{\mu}_1,\lambda_{\bbox{\epsilon}_1}'|\frac{1}{\cos\alpha}
|\Lambda,\bbox{\mu}_2,\lambda_{\bbox{\epsilon}_1}\rangle,\label{VC3body}
\end{eqnarray}
with the further simplification $\gamma=-\beta$. The double
summation (transforming the bra and ket vectors between Jacobi trees) 
seems to spoil the present approach's advantage over a multipole expansion.
After all, the latter also leads to two summations, namely, a sum over the 
multipole order and another summation stemming from the analytical evaluation 
of the nested integral over $\alpha$. Note, however, that the present method 
achieves significantly more with two summations: it accounts for {\em all the
cusps} in the wave functions whenever an inter-particle separation vanishes.

Finally, the particular case of a three-body system involves two relevant
Jacobi vectors only. All possible Jacobi trees are thus related to one 
another by rotations in the {\em same plane} $(\bbox{\xi}_1,\bbox{\xi}_2)$.
This particularity of the three-body problem might suggest arranging the
Cartesian coordinates of the Jacobi vectors in the following way:
\begin{equation}
\{x_1,x_2,y_1,y_2,z_1,z_2\},\label{otherchoice}
\end{equation}
rather than our choice (\ref{arrangement}). The $H_j$ resulting from the
above arrangement of coordinates coincide with $J_{x_1x_2}$, $J_{y_1y_2}$,
and $J_{z_1z_2}$, thereby simplifying the rotation of harmonics: 
{\em All} rotations reduce to the first case of $(z_1z_2)$-type rotations
discussed in Sec.~\ref{rotharms}, with matrix elements
\begin{equation}
\langle\Lambda',\bbox{\mu}',\lambda_{\bbox{\epsilon}_1}'|
\exp(i\beta[J_{x_1x_2}+J_{y_1y_2}+J_{z_1z_2}])|\Lambda,\bbox{\mu},\lambda_{
\bbox{\epsilon}_1}\rangle = e^{i(m_1+m_2+m_3)\beta}\delta_{\Lambda'\Lambda}
\delta_{\bbox{\mu}'\bbox{\mu}}\delta_{\lambda_{\bbox{\epsilon}_1}'\lambda_{
\bbox{\epsilon}_1}}. 
\end{equation}
This orthogonality relation virtually eliminates the double 
summation over $\bbox{\mu}_1,\bbox{\mu}_2$ in (\ref{VC3body}).
However, the corresponding harmonics are now
functions of $(x_1\pm ix_2)$, $(y_1\pm iy_2)$, and $(z_1\pm iz_2)$. 
Transforming to hyperspherical coordinates using 
\begin{equation}\label{otherHScoord}
\begin{array}{rcl}
x_1 & = & R\sin\alpha_1\cos\alpha_2\cos\varphi_1, \\
y_1 & = & R\sin\alpha_1\sin\alpha_2\cos\varphi_2, \\
z_1 & = & R\cos\alpha_1\cos\varphi_3, 
\end{array}
\qquad\qquad
\begin{array}{rcl}
x_2 & = & R\sin\alpha_1\cos\alpha_2\sin\varphi_1, \\
y_2 & = & R\sin\alpha_1\sin\alpha_2\sin\varphi_2, \\
z_2 & = & R\cos\alpha_1\sin\varphi_3, 
\end{array}
\end{equation}
yields $H_j=-i\frac{\partial}{\partial\varphi_j}$, 
with $0\le\varphi_j\le 2\pi$ and $0\le\alpha_k\le\pi/2$. 
While the $H_j$, as well as the harmonics,  now obviously
attain their simplest form, the choice (\ref{otherchoice}) has two serious 
drawbacks: (i) None of the three eigenvalues $m_j$ of the $H_j$ have
physical significance, and (ii) $1/|\bbox{\xi}_1|=(x_1^2+y_1^2+z_1^2)^{-1/2}$ 
is a (complicated) function of all five angles 
$(\alpha_1,\alpha_2,\varphi_1,\varphi_2,\varphi_3)$. The latter problem is
solved by using hyperspherical coordinates (\ref{HScoords}) instead
of (\ref{otherHScoord}), yielding for each of the combinations $(x_1\pm ix_2)$,
$(y_1\pm iy_2)$, and $(z_1\pm iz_2)$ a sum of two terms (as opposed to the
single terms obtained for $(x_j\pm iy_j)$ in the previous sub-sections). 
Expanding powers of these binomials leads to two additional summations, 
leaving us with no net gain.
\section{Conclusion}
\label{conclusion}
Recognizing the independent-particle model's failure to account for cusps
due to variables 
on which the wave function does not depend explicitly, we have developed
a method that satisfies Kato's cusp condition through reference frame
transformations. By transforming the {\em wave function} to the appropriate
reference frame, we expose the wave function's cusp arising from  vanishing
of any given particle separation $r_{ij}$. In addition to satisfying the cusp
condition on the wave function, this technique also simplifies the
calculation of the pairwise Coulomb interaction $\sim 1/r_{ij}$, as compared
to the conventional multipole expansion.

Implementation of the approach outlined above resolves into three major
tasks, all addressed in the present investigation: (i) the systematic study 
of the
relevant transformations between reference frames; (ii) the definition of
functions suitable for such transformations; (iii) the determination of the
transformation matrix. While the literature on
Lie algebra provides ready-made solutions to problem (i), it usually fails
to do so for (ii) and (iii). Furthermore, the mathematical literature does
not exploit the particularity of an atomic or molecular $N$-body system.

More specifically, problem (i) is solved by using mass-scaled Jacobi 
coordinates, since the transformations between reference frames reduce then
to generalized rotations in $(3N-3)$ dimensions. The Lie algebra $so(3N-3)$
describes these transformations completely, embodied in the sets of
commuting rotation operators $\{H_j\}$ and ladder operators
$\{E_{\pm\bbox{\epsilon}_j}\}$, $j=1,\ldots,\lfloor (3N-3)/2\rfloor$.
To solve problem (ii) mentioned above, we have introduced basis functions
defined entirely through their behavior under infinitesimal rotations, i.e.,
when acted upon by the operators $H_j$ and
$E_{\pm\bbox{\epsilon}_j}$. Simultaneous eigenfunctions of all the $H_j$
constitute appropriate basis functions, classified further according to
their matrix elements for the transformation between reference frames. The
latter step removes possible ambiguities whenever the $H_j$ have degenerate
eigenvalues. Finally, by extending the concept of Euler-angle rotations from
three to higher dimensions, we have provided the solution to task (iii), the
determination of the transformation matrix elements. For an arbitrary
$N$-body system, each change of reference frames considered here resolves
into a sequence of basic transformations described by these matrix elements.

In an $N$-body system, the high dimensionality
arises from the {\em product structure} of the $(N-1)$-dimensional
particle-space and the three-dimensional physical space. The basic step in the
transformation between reference frames reduces to a two-dimensional
rotation in a plane of particle-space. Upon expansion of the particle-space
variables into their physical-space components, the basic rotation induces
three two-dimensional rotations in $(3N-3)$-dimensional space that are
analyzed using the Lie algebra $so(3N-3)$. To take
advantage of the efficiency offered by the Lie algebraic method, we treat
the transformations as rotations in a genuine $(3N-3)$-dimensional space.
Thus, we work with the {\em first-order} differential operators $H_j$ and
$E_{\pm\bbox{\epsilon}_j}$ throughout. 
Arranging the generic Cartesian components $(x_j,y_j,z_j)$ of the  
particle-space variables $\bbox{\xi}_j$ appropriately, we can 
\emph{partially} recover the product structure characteristic
for the physical application at hand: The eigenvalues of $(N-1)$ among the
$H_j$ carry physical significance and we interpret them as $z$-projections
of orbital angular momenta. Their sum represents the total angular momentum's
$z$-projection, an invariant of the system. However, neither the
individual angular momenta, nor the coupled (total) angular momentum can
appear in our treatment, as they are represented by second-order operators.
Other methods exploit the product structure of the $(3N-3)$-dimensional 
coordinate space to a larger extent by solving the Laplacian's (second-order) 
eigenvalue problem through separation of
variables. However, the resulting hyperspherical harmonics incorporate these
features of the three-dimensional physical 
space at considerable cost: Transforming these functions between reference
frames proves very inefficient, requiring essentially their expansion into the
equivalent sets of harmonics introduced in the present investigation.

In conclusion, we have demonstrated by explicit construction the possibility 
of describing a system of $N$ charged particles 
without recourse to multipole expansions. Kato's cusp condition is satisfied
implicitly through reference frame transformations of the {\em wave function}. 
We introduced appropriate basis functions and discussed their symmetries under
particle interchange and reflection through the origin.
We derived the matrix elements for the transformation between reference frames 
for arbitrary numbers of particles, and we showed that the interaction 
matrix elements attain a simpler form as compared to the conventional multipole
expansion. Because the formalism underlying our approach relies exclusively
on first-order differential operators, it does not incorporate 
angular momenta, thereby limiting its usefulness for 
bound state problems. However, the technique introduced here provides 
significant simplifications 
in scattering problems where partial angular momenta are not resolved. 


\begin{thebibliography}{99}
\bibitem{irrep} The procedure outlined in the text generates only weights
of real, symmetric representations. As these representations account for
all sets of harmonics needed in our applications, 
irreducible representations of $so(d)$ with highest weight
other than $(\Lambda,0,\ldots,0)$ are not considered here.
\end{thebibliography}
\end{document}